\documentclass{article}
\usepackage[english]{babel}

\usepackage[letterpaper,top=2cm,bottom=2cm,left=3cm,right=3cm,marginparwidth=1.75cm]{geometry}

\usepackage{amsmath,amssymb}
\usepackage{graphicx}
\usepackage{fancyhdr}
\usepackage[colorlinks=true, allcolors=blue]{hyperref}
\usepackage{verbatim}
\usepackage[caption=false]{subfig}
\usepackage{authblk}

\begin{document}
\title{Estimating the master stability function from the time series of one oscillator via reservoir computing}
\author{Joseph D. Hart}% \textsuperscript{*}}
%\email{joseph.hart@nrl.navy.mil}
%\affiliation{
 \affil{
 US Naval Research Laboratory, Code 5675, Washington, DC, USA 20375 \\ *joseph.hart@nrl.navy.mil
}

\maketitle

\begin{abstract}
The master stability function (MSF) yields the stability of the globally synchronized state of a network of identical oscillators in terms of the eigenvalues of the adjacency matrix. In order to compute the MSF, one must have an accurate model of an uncoupled oscillator, but often such a model does not exist. We present a reservoir computing technique for estimating the MSF given only the time series of a single, uncoupled oscillator. We demonstrate the generality of our technique by considering a variety of coupling configurations of networks consisting of Lorenz oscillators or H{\'e}non maps.
\end{abstract}
%\maketitle

Synchronization is an emergent behavior in which a number of interacting oscillators do the same thing at the same time \cite{pikovsky2001universal}. Synchronization is a crucial phenomenon in various systems, such as power grids \cite{motter2013spontaneous}, neuronal networks \cite{dhamala2004enhancement}, cardiac tissue \cite{ji2017synchronization},  and chemical \cite{taylor2009dynamical}, electronic \cite{pecora1990synchronization},  opto-electronic \cite{ravoori2011robustness,hart2015adding}, and laser \cite{kinzel2009synchronization} systems.  When the oscillators are discrete objects, the interactions can be described by a network, and so the synchronization of oscillator networks has long been a rich area of research \cite{pikovsky2001universal,lindsey1985network,barahona2002synchronization}. 

The problem of global synchronization and its stability was solved by the Master Stability Function (MSF) approach \cite{pecora1998master}. Once determined, the MSF gives the stability of the globally synchronized state of a network of identical oscillators for any network topology from the eigenvalues of the network Laplacian adjacency matrix. 

%In order to obtain the MSF, one must have either an accurate model of an uncoupled oscillator and the coupling function \cite{pecora1998master} or a set of three coupled oscillators with controllable coupling strengths \cite{fink2000three}. In many situations, one may not have a model, and building such a tunable network may be difficult; yet one may still desire to know the stability properties of network synchronization. In this work, we present a machine learning technique for estimating the MSF given only the time series of a single, uncoupled oscillator. We also find that the machine-learned model can provide an accurate estimate of the MSF even when it is not successful at attractor reconstruction.

In order to obtain the MSF, one must have either an accurate model of an uncoupled oscillator and the coupling function \cite{pecora1998master} or a set of three coupled oscillators with controllable coupling strengths \cite{fink2000three}. In many situations, one may not have a model, and building such a tunable network may be difficult; yet one may still desire to know the stability properties of network synchronization. In this work, we demonstrate that reservoir computing can be used as a machine learning technique for estimating the MSF given only the time series of a single, uncoupled oscillator. We also find that the trained reservoir computer can provide an accurate estimate of the MSF even when it is not successful at attractor reconstruction.

In our technique, we train a reservoir computer to reproduce the dynamics of a single oscillator for which we have the time series; this task is often called ``attractor reconstruction,'' and can be performed with a variety of machine learning modalities \cite{pathak2017using,itoh2017reconstructing,lu2018attractor,vlachas2020backpropagation,gauthier2021next}. The trained reservoir computer now serves as a model for the uncoupled oscillator. If one knows the form of the coupling, one can compute the MSF of the learned model and use this as an estimate of the MSF of the true oscillator system. We find that the MSF of the reservoir computer can provide an excellent estimate of the MSF of the true oscillator system, for both a continuous-time system (the Lorenz system) and a discrete-time system (the H{\'e}non map, see Supplemental Material). %This is a remarkable result: The machine-learned model 

The trained reservoir computer is a high-dimensional dynamical system that displays not only the trained dynamical behaviors of the true oscillator, but also the untrained networked dynamical behaviors of the true oscillator near the synchronization manifold. While previous works have shown that an accurate estimate of the Lyapunov spectrum of a single oscillator can be obtained from a reservoir computer that achieves attractor reconstruction \cite{pathak2017using}, in this work we demonstrate that a well-trained reservoir computer can provide an accurate estimate of the MSF (i.e., the largest transverse Lyapunov exponent of the master stability equation). These are distinct tasks: In the Supplemental Material, we show that successful attractor reconstruction is not necessary for obtaining an accurate estimate of the MSF. 

In order for the MSF estimation to be accurate, the largest (of many) of the transverse Lyapunov exponents of the reservoir computer master stability equation must be approximately equal to the largest transverse Lyapunov exponent of the true master stability equation. We find that this behavior is more likely to occur when the spectral radius of the reservoir adjacency matrix is significantly less than one, a parameter regime in which reservoir computers are rarely operated, and we provide a reason why a small spectral radius is preferable.

While the dynamics of coupled reservoir computers have been considered previously \cite{hu2022synchronization,weng2023synchronization}, no correspondence has been shown between the reservoir computer coupling strength and the true oscillator coupling strength. In this work, we present a novel coupling scheme in which the coupling strength is proportional to the sampling time of the training data $\tau$. We demonstrate that this factor of $\tau$ is essential for obtaining the correct MSF.

\section{Networks of Coupled Oscillators}
Consider a single oscillator with isolated (uncoupled) dynamics described by
\begin{equation}
\label{eq:UncoupledFlow}
    \dot{\mathbf{x}}(t) = \mathbf{F}(\mathbf{x}(t)),
\end{equation}
where $\mathbf{x}\equiv[x^{(1)},...,x^{(\mathcal{D})}]$ is the state vector of a single node. One can couple a network of such oscillators together as follows:
\begin{equation}
\label{eq:CoupledFlow}
    \dot{\mathbf{x}}_i(t) = \mathbf \mathbf{F}(\mathbf{x}_i(t)) - \epsilon L_{ij}\mathbf{H}(\mathbf{x}_j(t))
\end{equation}
where $H$ is a function that describes the coupling and the sum over the subscript $j$ is implied. In this work we are concerned with the stability of the globally synchronized state $\mathbf{x}_i(t)=\mathbf{x}_j(t)$ for all pairs $i$ and $j$ as $t\to\infty$. The existence of the synchronized state is guaranteed by the Laplacian adjacency matrix $L$ \cite{pecora1998master}. The stability of the synchronized state is determined by the MSF \cite{pecora1998master}. 

\section{Reservoir Computing}

In this work, we choose a reservoir computer \cite{lukovsevivcius2009reservoir,lukovsevivcius2012reservoir} (also called an echo state network \cite{jaeger2004harnessing}) as our machine learning modality. A reservoir computer is a type of recurrent neural network that is designed to be particularly easy to train and has been shown to be well-suited to the performance of a variety of time-series tasks, including attractor reconstruction \cite{pathak2017using,lu2018attractor,rohm2021model,kong2023reservoir}, Lyapunov exponent estimation \cite{pathak2017using,lu2018attractor}, causal inference \cite{banerjee2019using,banerjee2021machine}, and nonlinear control \cite{canaday2021model}. A reservoir computer has also been shown to have the ability to predict un-trained bifurcations \cite{rohm2021model,kong2023reservoir},  to forecast network dynamics \cite{srinivasan2022parallel}, and to synchronize with either the true chaotic system \cite{weng2019synchronization} or with identical copies of itself \cite{fan2021anticipating,hu2022synchronization,weng2023synchronization}. We use a reservoir computer for these reasons, but we note that MSF estimation and can be tried with other machine learning modalities such as LSTM \cite{vlachas2020backpropagation}, NVAR \cite{gauthier2021next}, and extreme learning machines \cite{huang2006extreme}.
%A reservoir computer is a type of recurrent neural network that is computationally efficient to train and has been shown to be well-suited to the replication of nonlinear dynamical systems. 

Following Ref. \cite{pathak2017using}, the reservoir computers used here consist of a randomly-designed recurrent neural network of $N$ discrete-time nodes. Let $\mathbf{r}[n]$ be an $N\times 1$ column vector that describes the state of the reservoir at time $n$. The reservoir computer is described in training mode by

\begin{equation}
\mathbf{r}[n+1] = 
\tanh{(A\mathbf{r}[n]+W^{in}\mathbf{x}[n]+\mathbf{b})},
\end{equation}
where $\mathbf{x}[n]$ is an $\mathcal{D}\times 1$ column vector describing the input signal to the reservoir, $W^{in}$ is an $N\times\mathcal{D}$ matrix describing the input layer, $A$ is an $N\times N$ matrix describing the inter-nodal connections in the reservoir layer, and $\mathbf{b}$ is an $N\times 1$ bias vector, the purpose of which is to break the symmetry of the hyperbolic tangent function. In this work, we consider input signals that are created by sampling a continuous-time dynamical system described by the state vector $\mathbf{x}(t)$, such that $\mathbf{x}[n]\equiv\mathbf{x}(n\tau)$ where $\tau$ is the uniform sampling time.

In the training phase the reservoir is driven by the input signal $\mathbf{x}[n]$ and is trained to predict that signal at the next time step $\mathbf{x}[n+1]$; this prediction $\hat{\mathbf{x}}$ is obtained from the reservoir by

\begin{equation}
\label{eq:xhat}
\hat{\mathbf{x}}[n] = W^{out}\mathbf{P}(\mathbf{r}[n]),    
\end{equation}
where $\mathbf{P}$ is an $N_p$-dimensional function of $\mathbf{r}$, and $W^{out}$ is a $\mathcal{D}\times N_p$ matrix obtained by ridge regression \cite{tikhonov1995numerical}. Often, the choice $P(\mathbf{x})=\mathbf{x}$ is made.

Once the training is complete, one can turn the reservoir into an autonomous dynamical system by feeding this prediction back into the reservoir according to:
\begin{equation} \begin{split}
\label{eq:autonomous_RC}
\mathbf{r}[n+1] &= \tanh{(A\mathbf{r}[n]+W^{in}\hat{\mathbf{x}}[n]+\mathbf{b})}.
\end{split} \end{equation}

It is known that, when the training succeeds, the autonomous reservoir computer described by Eq. \ref{eq:autonomous_RC} can provide a stable reconstruction of the attractor of a dynamical system \cite{pathak2017using,lu2018attractor,rohm2021model}. When the attractor is successfully reconstructed, short-term predictions are of course limited to a few Lyapunov times; however, the long-term climate of the true system is reproduced by the autonomous reservoir, and the Lyapunov exponents of Eq. \ref{eq:autonomous_RC} often agree with the Lyapunov exponents of the true dynamical system. 

%\subsection{Coupled reservoir computers}

Since an autonomous reservoir computer is just a dynamical system, a set of identical autonomous reservoir computers can be coupled together in a network. Let $\mathbf{r}_i$ represent the $N\times 1$ state vector of the $i^{th}$ reservoir. We hypothesize that it will be interesting and useful to couple together $M$ autonomous reservoir computers in the following way

\begin{equation} \begin{split}
\label{eq:CoupledReservoirs1}
    \mathbf{r}_i[n+1]& = \tanh\big(A\mathbf{r}_i[n]\\&+W^{in}\{\mathbf{\hat{x}}_i[n]-\epsilon\tau L_{ij} \mathbf{H}(\mathbf{\hat{x}}_j[n])\}+\mathbf{b}\big)
\end{split} \end{equation}
where the sum over $j$ is implied, $L$ is the $M\times M$ Laplacian coupling matrix, $\epsilon$ is an overall coupling strength, and $\mathbf{H}$ is a coupling function that describes how and which elements of the output of reservoir $j$ couple into which input elements of reservoir $i$. We hypothesize that this is analogous to Eq. \ref{eq:CoupledFlow}.  While coupled reservoir computers have been considered previously using different coupling schemes, in the coupling scheme used here the coupling strength is proportional to the sampling time $\tau$ of the training data. Some justification for this is presented in the Supplemental Material. As we show, this results in the equivalence of the coupling strength $\epsilon$ in the ordinary differential equation Eq. \ref{eq:CoupledFlow} and in the discrete-time reservoir equation Eq. \ref{eq:CoupledReservoirs1} when $\tau\ll 1/\epsilon\lambda_k$, where $\lambda_k$ are the eigenvalues of $L$.

We assume that the form of the coupling $\mathbf{H}(\mathbf{x})$ is known. In cases where it is not known, there are a variety of techniques for estimating the coupling function \cite{stankovski2017coupling}.

%Equation \ref{eq:CoupledReservoirs1} describes the dynamics of a set of identical reservoir computers coupled in a network described by the Laplacian matrix $L.$ The structure of the coupling permits the autonomous reservoir computers to display global synchronization. However, an important question is: \textit{Can a network of coupled identical reservoir computers with a given topology synchronize stably?} This question can be answered by computing the MSF \cite{pecora1998master} of the autonomous reservoir computer. 

\section{Master Stability Function for coupled reservoir computers}

To compute the MSF, one needs both the dynamics of the synchronized state (Eq. \ref{eq:UncoupledFlow}) and the Jacobian of the variational equation corresponding to Eq. \ref{eq:CoupledFlow} in the basis in which $L$ is diagonal. The former is provided by the trained reservoir Eq \ref{eq:autonomous_RC}. Alternatively, it could be provided the measured dynamics of the nonlinear oscillator, as we demonstrate the Supplemental Material.

To estimate the latter we linearize Eq. \ref{eq:CoupledReservoirs1} about the synchronous flow given by Eq. \ref{eq:autonomous_RC}, where we use a subscript $s$ to indicate the synchronized state:

\begin{equation}
\begin{split}
\delta\mathbf{r}_i[n+1] = &\mathrm{sech}^2\big(A\mathbf{r}_s[n]+W^{in}\hat{\mathbf{x}}_s[n]+\mathbf{b}\big)\\ &\big\{A\delta\mathbf{r}_i[n] + W^{in}\delta\hat{\mathbf{x}}_i[n]\big\} \\ 
\delta\hat{\mathbf{x}}_i[n] =& W^{out}d\mathbf{P}(\mathbf{r}_s[n])\delta\mathbf{r}_i \\- \epsilon\tau L_{ij}&d\mathbf{H}(\hat{\mathbf{x}}_s[n])W^{out}d\mathbf{P}(\mathbf{r}_s[n])\delta\mathbf{r}_j[n].
    \end{split}
 \end{equation}

We want to consider only variations that are transverse to the synchronization manifold. The first term in the curly brackets is block diagonal with $M\times M$ blocks. The second term is treated by performing a linear transformation to a basis in which $L$ is diagonal. This transformation does not affect the first term. This results in the master stability equation:

\begin{equation}
\label{eq:MasterStabilityEquation}
\begin{split}
    \delta\mathbf{q}&_k[n+1] = \mathrm{sech}^2\big(A\mathbf{r}_s[n]+W^{in}\hat{\mathbf{x}}_s[n]+\mathbf{b}\big)D \\ 
D&=\big\{A + W^{in} W^{out}d\mathbf{P}(\mathbf{r}_s[n]) \\&- \epsilon\tau \lambda_k W^{in}d\mathbf{H}(\hat{\mathbf{x}}_s[n])W^{out}d\mathbf{P}(\mathbf{r}_s[n])\big\}\delta\mathbf{q}_k[n].
    \end{split}
\end{equation}
where we define $K\equiv\epsilon\lambda$ and $\lambda_k$ are the $M$ eigenvalues of $L$. For $K=\epsilon\lambda_k=0$, Eq. \ref{eq:MasterStabilityEquation} is the variational equation for the synchronization manifold. All other eigenvalues correspond to transverse eigenvectors and therefore determine the stability of the globally synchronized state.

The MSF is the largest Lyapunov exponent of Eq. \ref{eq:MasterStabilityEquation} as a function of the complex number $K$. Once computed, the sign of the MSF gives the stability of global synchronization of any reservoir computer network described by any Laplacian coupling matrix $L$ in terms of the eigenvalues of $L$. 

\section{Results}

\begin{figure*}[htb]

\subfloat[]{%
  \includegraphics[width=0.33\textwidth]{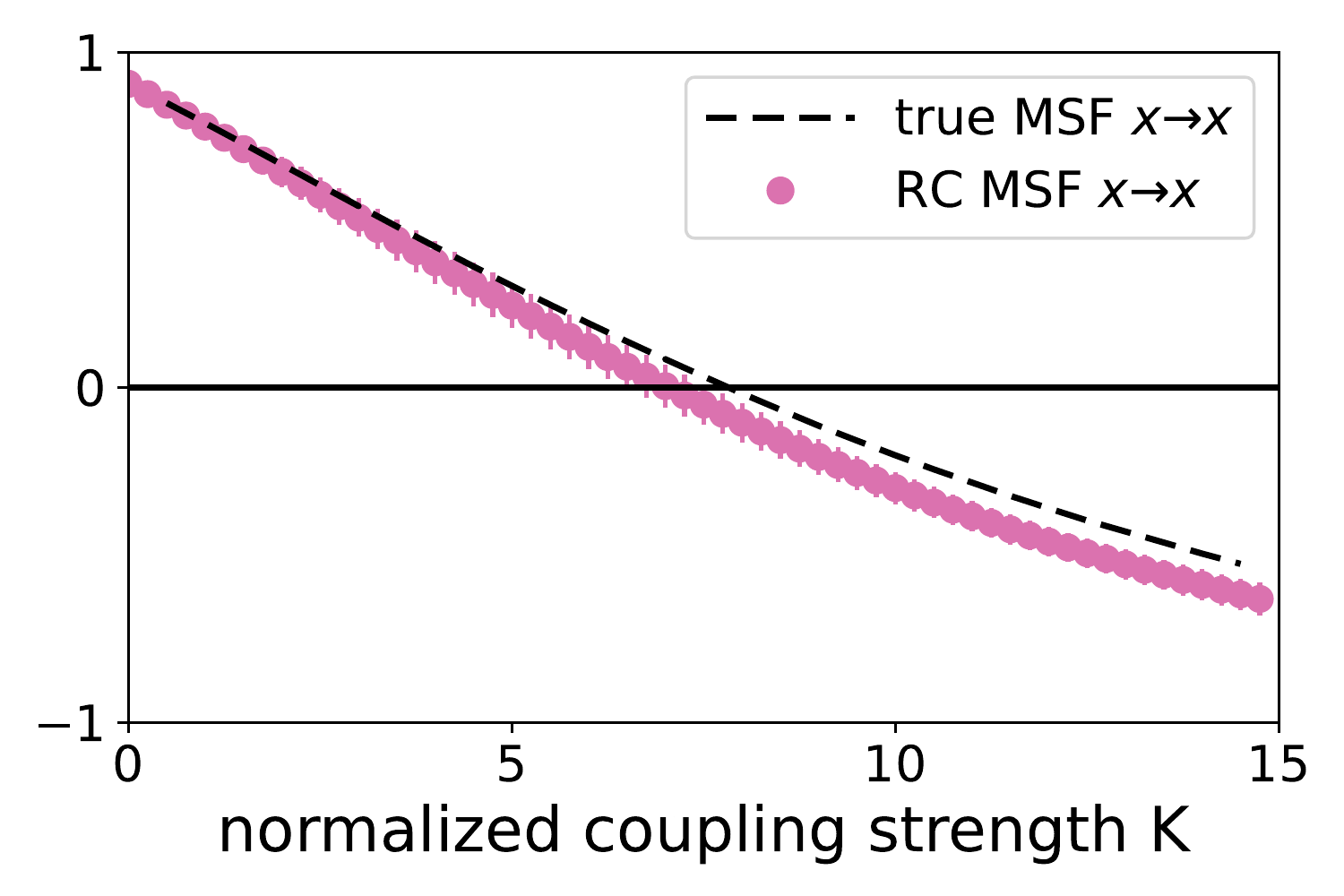}%
}\hfill
\subfloat[]{%
  \includegraphics[width=0.33\textwidth]{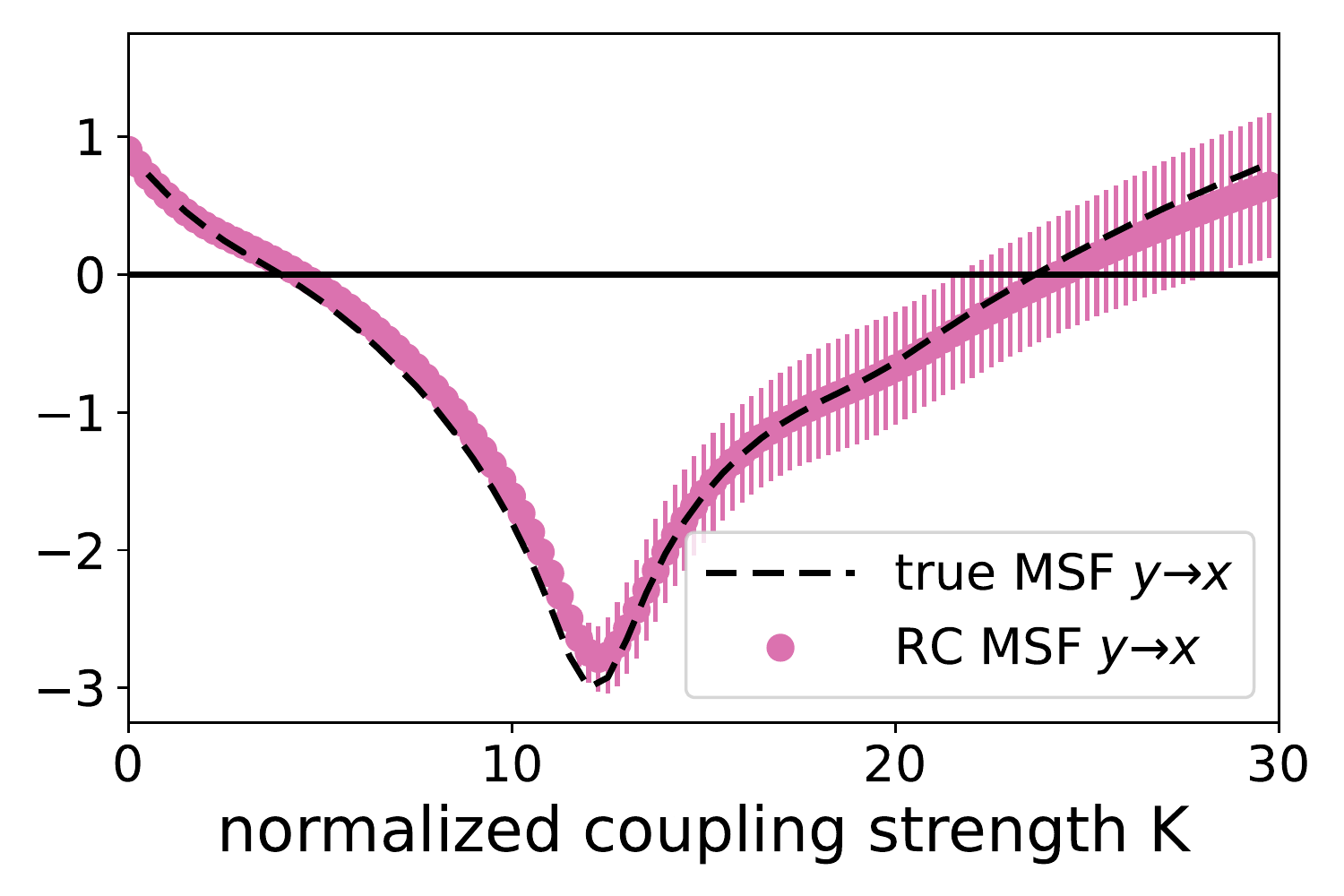}%
}\hfill
\subfloat[]{%
  \includegraphics[width=0.33\textwidth]{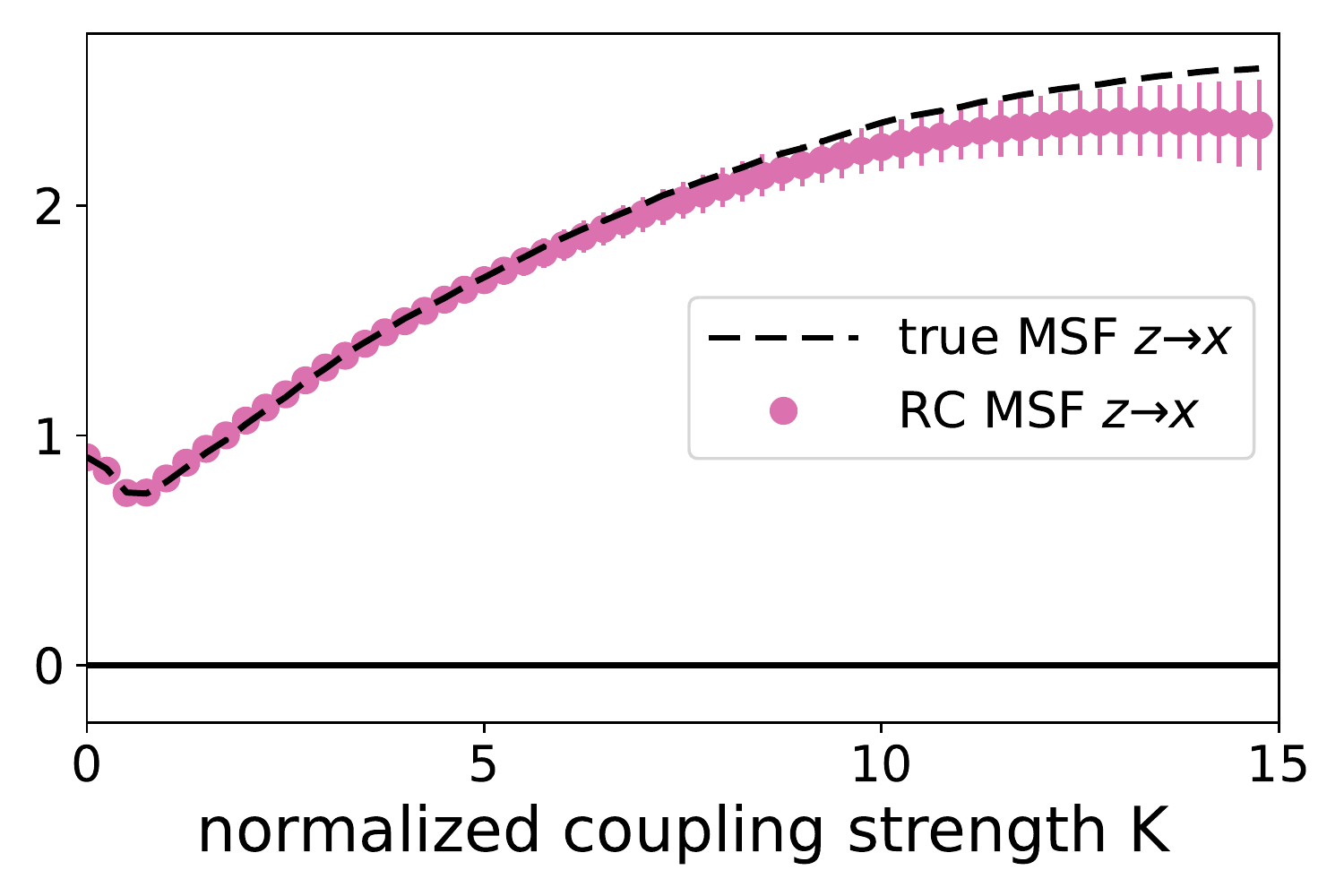}%
}\hfill

\subfloat[]{%
  \includegraphics[width=0.33\textwidth]{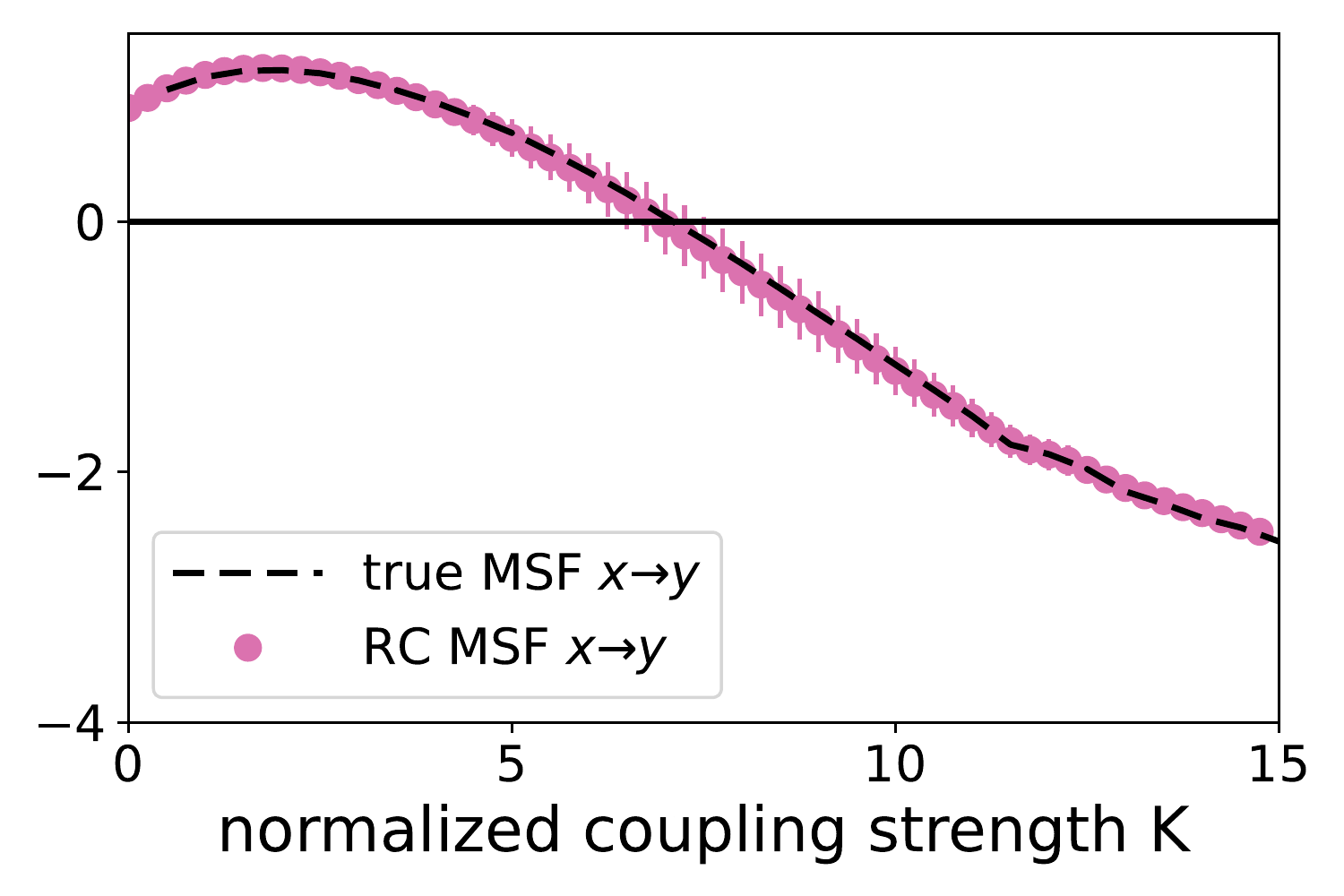}%
}\hfill
\subfloat[]{%
  \includegraphics[width=0.33\textwidth]{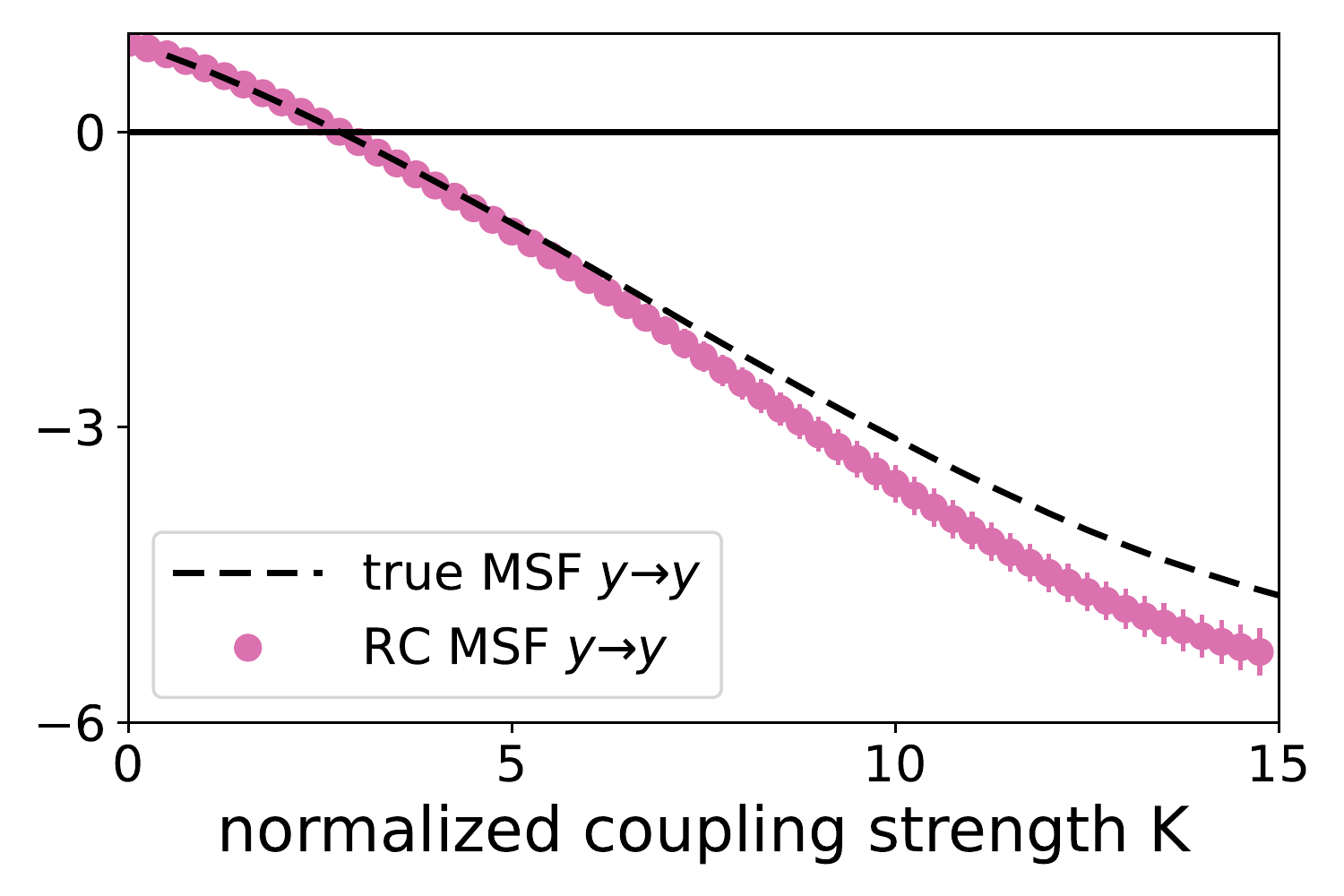}%
}\hfill
\subfloat[]{%
  \includegraphics[width=0.33\textwidth]{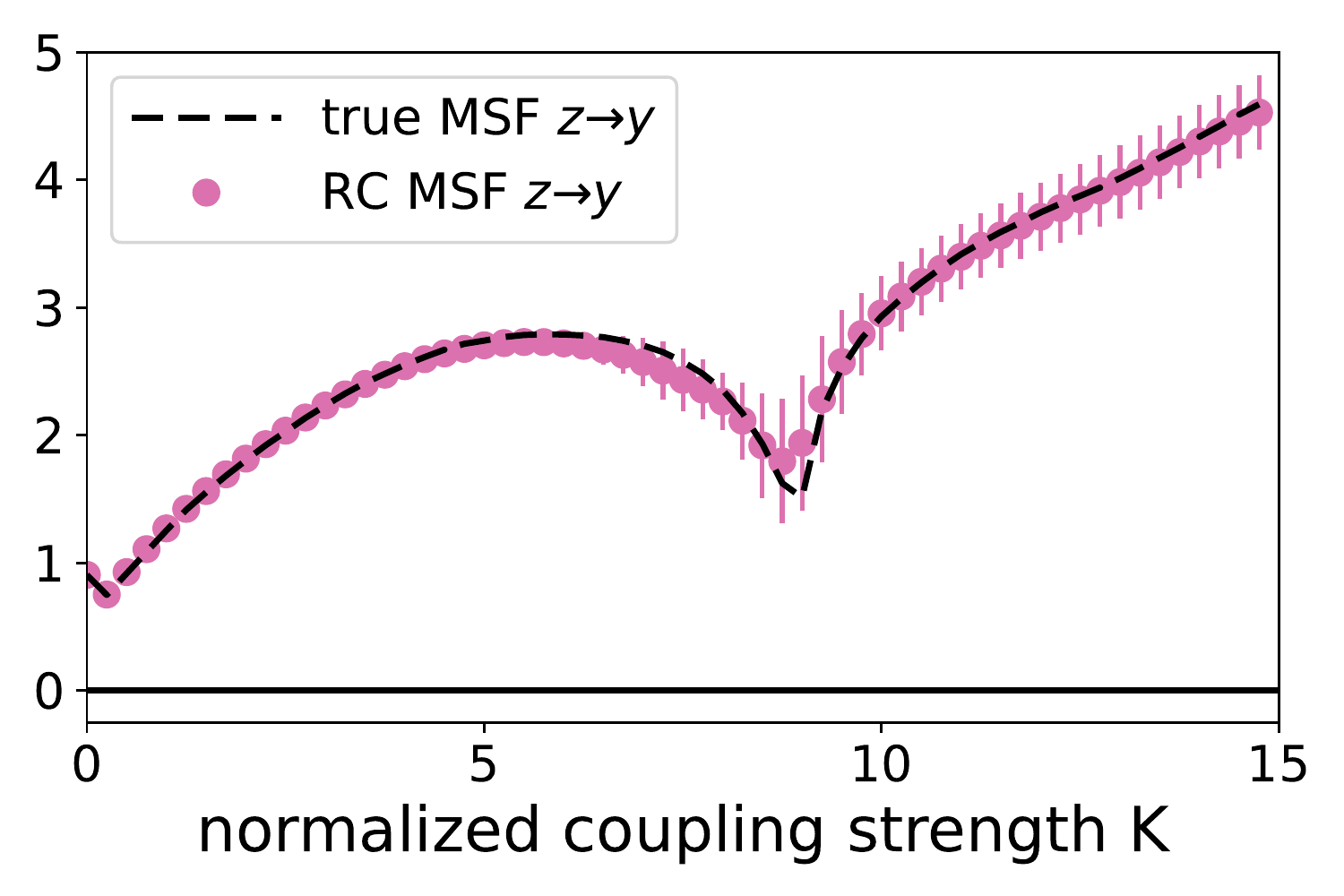}%
}\hfill

\subfloat[]{%
  \includegraphics[width=0.33\textwidth]{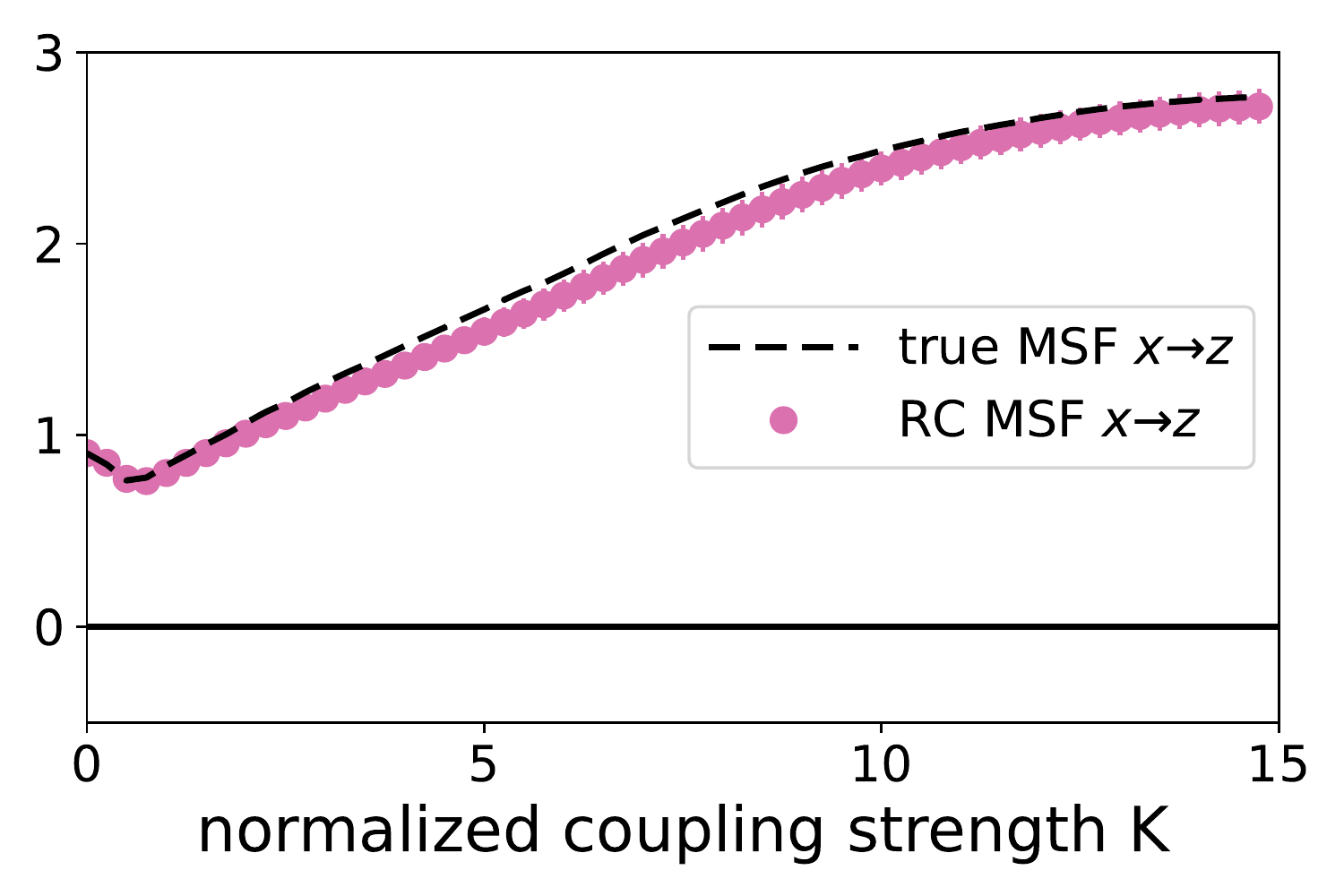}%
}\hfill
\subfloat[]{%
  \includegraphics[width=0.33\textwidth]{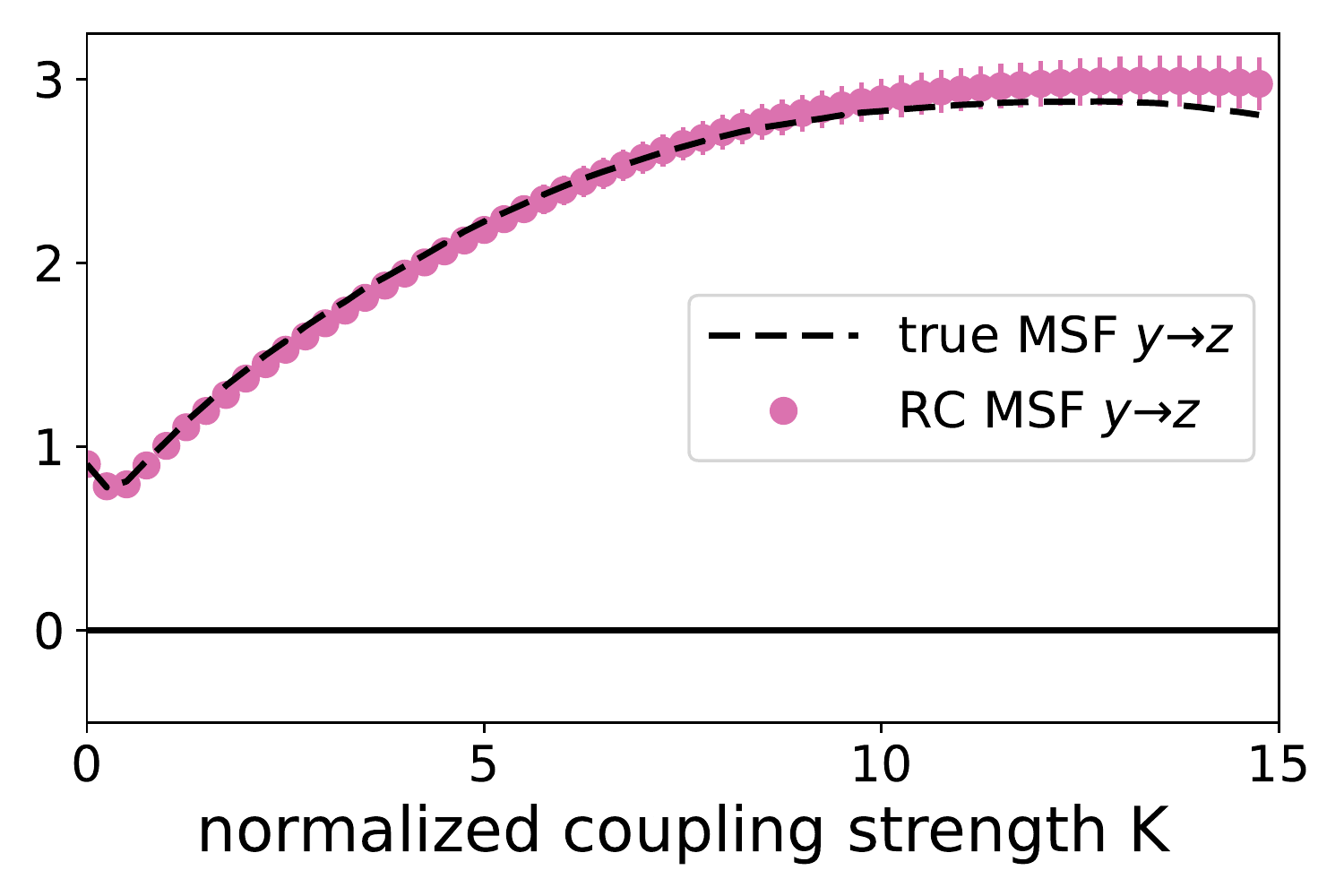}%
}\hfill
\subfloat[]{%
  \includegraphics[width=0.33\textwidth]{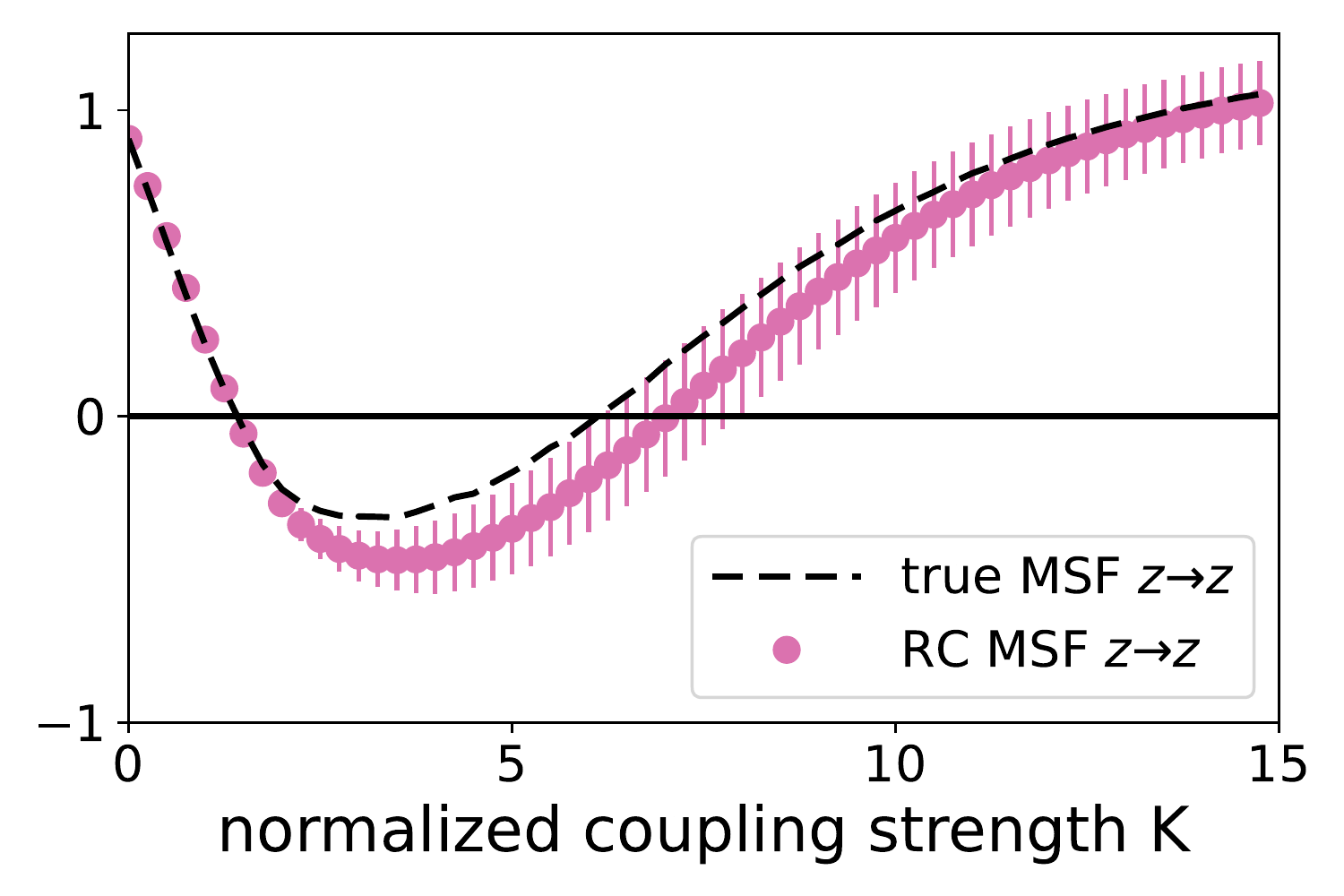}%
}\hfill
\caption{Master Stability Function of the true Lorenz system (black dashed line) and the trained reservoir computer (pink dots). A solid black line is shown at 0 to indicate the stability boundary.}
    \label{fig:MSF}
\end{figure*}
\begin{comment}
\begin{figure*}[ht]
    \centering
    \begin{subfigure}{0.31\textwidth}    \includegraphics[width=\textwidth]{MSF_Lorenz_x_to_x_Coupling_2.pdf}
    \end{subfigure}
    \begin{subfigure}{0.31\textwidth}    \includegraphics[width=\textwidth]{MSF_Lorenz_y_to_x_Coupling_2.pdf}
    \end{subfigure}
    \begin{subfigure}{0.31\textwidth}    \includegraphics[width=\textwidth]{MSF_Lorenz_z_to_x_Coupling_2.pdf}
    \end{subfigure}
    \begin{subfigure}{0.31\textwidth}    \includegraphics[width=\textwidth]{MSF_Lorenz_x_to_y_Coupling_2.pdf}
    \end{subfigure}
    \begin{subfigure}{0.31\textwidth}    \includegraphics[width=\textwidth]{MSF_Lorenz_y_to_y_Coupling_2.pdf}
    \end{subfigure}
    \begin{subfigure}{0.31\textwidth}    \includegraphics[width=\textwidth]{MSF_Lorenz_z_to_y_Coupling_2.pdf}
    \end{subfigure}
    \begin{subfigure}{0.31\textwidth}    \includegraphics[width=\textwidth]{MSF_Lorenz_x_to_z_Coupling_2.pdf}
    \end{subfigure}
    \begin{subfigure}{0.31\textwidth}    \includegraphics[width=\textwidth]{MSF_Lorenz_y_to_z_Coupling_2.pdf}
    \end{subfigure}
     \begin{subfigure}{0.31\textwidth}   \includegraphics[width=\textwidth]{MSF_Lorenz_z_to_z_Coupling_2.pdf}
    \end{subfigure}
    \caption{Master Stability Function of the true Lorenz system (black dashed line) and the coupled reservoir system (pink dots). A solid black line is shown at 0 to indicate the stability boundary.}
    \label{fig:MSF}
\end{figure*}
\end{comment}

\textit{Is the stability of synchronization of a network of coupled, trained reservoirs predictive of the stability of synchronization of a network of the true oscillators?} We investigate this question by training a reservoir computer on the Lorenz system with linear, one-component coupling \cite{huang2009generic} (e.g., for $x\to x$ coupling, $[\mathbf{H}(\mathbf{x}))]_{ij}=x$ for $i=j=1$ and zero otherwise), and comparing the MSF of the coupled reservoir system with the MSF of the true Lorenz system. In the Supplemental Material, we also demonstrate that this technique can be used to estimate the MSF of the R{\"o}ssler oscillator and the discrete-time H{\'e}non map.

We use the same type of reservoir as in Ref. \cite{pathak2017using}. The 300-node reservoir was set up as follows. To form the adjacency matrix, a sparse random Erdos-Renyi network with average degree 6 was created, with each non-zero element of $\tilde{A}$ was drawn independently and uniformly from the range [-1, 1]. All elements were then scaled such that the spectral radius is $\rho$. The input matrix $W^{in}$ was size $300\times 3$. The first 100 elements of the first column, the second 100 elements of the second column, and the final 100 elements of the third column were non-zero; the remaining elements were zero. The non-zero elements of $W^{in}$ were randomly drawn independently and uniformly from $[-\sigma, \sigma]$. No bias was used ($\mathbf{b}=0$).

Following Ref. \cite{pathak2017using}, the reservoir output $\hat{\mathbf{x}}=[\hat{x},\hat{y},\hat{z}]$ is given by

\begin{equation}
    \begin{bmatrix}
        \hat{x}[n] \\ \hat{y}[n] \\ \hat{z}[n]
    \end{bmatrix} = W^{out}\begin{bmatrix}\mathbf{r}[n] \\
        \mathbf{r}[n] \\
        \tilde{\mathbf{r}}[n] 
    \end{bmatrix},
\end{equation}
where $\tilde{\mathbf{r}}$ is defined such that  the first half of its elements are the same as that of $\mathbf{r}$, while $\tilde{\mathbf{r}}= r^2$ for the remaining half of the reservoir nodes. The $W^{out}$ matrix is trained by ridge regression. 

The input Lorenz signal used was obtained by integrating the Lorenz equations \cite{lorenz1963deterministic} using a 2nd-order Runge-Kutta method with time step $dt=0.001$, then downsampling the signal to have sampling time $\tau=0.02$. Dynamical noise was by modeled by adding white Gaussian noise of standard deviation 0.01$\sqrt{dt}$ at each time step according to the method described in Ref. \cite{honeycutt1992stochastic}; the presence of dynamical noise seems to be beneficial, consistent with previous work on network reconstruction using reservoir computing \cite{banerjee2019using,banerjee2021machine}. We found that a ridge parameter of 0 worked best, perhaps because the noise was sufficient to prevent overfitting \cite{bishop1995training,wikner2022stabilizing}. The Lyapunov exponents of Eq. \ref{eq:MasterStabilityEquation} were computed using the QR method \cite{geist1990comparison}.

We trained 1000 different Erd{\"o}s-Renyi reservoirs with spectral radius $\rho$ drawn from a uniform distribution with limits [0,0.3] and $\sigma$ from a uniform distribution with limits [0,0.2] on the same Lorenz time series. We note that these values of $\rho$ are much smaller than that typically advised in the literature; one reason for why small values of $\rho$ may be better in this case is provided in the Supplemental Material. We used 40000 training time steps. We then computed the MSF of the 100 reservoirs that gave the best prediction error over seven Lyapunov times. Each MSF was computed over 50000 time steps. MSFs that were extreme outliers relative to the population were discarded; the rest were retained. In all cases, more than 90 MSFs were retained. The results shown are the mean MSF of the retained reservoirs, and the error bars are the standard deviation. The exclusion of outlier MSFs had little effect on the mean, but did reduce the standard deviation in some cases.

We first tested whether our reservoirs could reproduce the Lyapunov exponents of the true Lorenz system (that is, we computed the Lyapunov exponents of Eq. \ref{eq:MasterStabilityEquation} with $K=0$). An autonomous reservoir has many Lyapunov exponents: for the three largest (averaged over all 100 retained reservoirs), we obtained: 0.906, 0.000, $-14.40$. The true Lorenz Lyapunov exponents are: 0.906, 0.000, $-14.6$. It is known that reservoir computers often struggle to replicate the negative Lyapunov exponent of the Lorenz system \cite{pathak2017using}, though it is interesting to note that our reservoir computer (with its small spectral radius) comes closer than the large spectral radius reservoir computer of Ref. \cite{pathak2017using}.

We then computed the MSF for real $K$ of the true Lorenz system and of the coupled reservoir system Eq. \ref{eq:MasterStabilityEquation} for all possible single-variable linear coupling 
schemes. The results are shown in Fig. \ref{fig:MSF}. The true MSF appears to agree excellently with that computed in Ref. \cite{huang2009generic}. For the reservoir computer MSF, the results (error bars) shown are the average (standard deviation) over the retained reservoirs. The agreement between the reservoir computer MSF and the true MSF is excellent everywhere for cross-coupling (e.g., $x\to y$ and $y\to x$). For self-coupling ($x\to x$, etc.) the agreement is good though not perfect. It it not clear why the results are not as good for the case of self-coupling. We also note that, in general, the accuracy of the MSF estimation seems to be better for smaller values of $K$. The reason is that our scheme for coupling the reservoir computers is equivalent to the true coupling scheme only when $\tau\ll 1/K$; this is explained further in the Supplemental Material.

We have investigated the accuracy of the estimation of the MSF by reservoir computing on networks of coupled H{\'e}non maps in the Supplemental Material. We find good agreement between the true and estimated MSFs when $\rho$ is small, and provide analysis that demonstrates that the disagreement at very negative values of the MSF is due to the non-zero spectral radius of the reservoir. We further demonstrate the generality of our MSF estimation technique by using the measured dynamics of the synchronized state $\mathbf{x}_s$ instead of $\hat{\mathbf{x}}_s$ in Eq. \ref{eq:MasterStabilityEquation} on a reservoir computer trained on the R{\"o}ssler system (for which it has been found that obtaining a reliable
attractor reconstruction using reservoir computing is more difficult than for the Lorenz system \cite{scully2021measuring}) in the Supplemental Material. Driving the reservoir with $\mathbf{x}_s$ is particularly useful when attractor reconstruction fails but one still requires an estimate of the MSF. These results confirm that successful attractor reconstruction is not necessary for accurate MSF (or Lyapunov exponent) estimation.

\section{Conclusions}

We showed that the master stability function of a reservoir computer trained on time series from a single, uncoupled nonlinear oscillator can provide an accurate estimate of the master stability function of the true oscillator. Therefore, a reservoir computer trained on time series from a single nonlinear oscillator can be used to determine the stability of any network of coupled oscillators of the same type. We demonstrated this on a variety of coupling configurations of the chaotic Lorenz oscillator and the chaotic H{\'e}non map (Supplemental Material), demonstrating that this technique can work for both continuous- and discrete-time oscillators.

The technique presented here offers the potential to solve the problem of global network synchronization by a simple measurement of an uncoupled oscillator. More generally, this result demonstrates that machine-learned systems, when coupled together, can quantitatively replicate the dynamical behavior of a true oscillator network, indicating a deep relationship between the trained reservoir computer and true oscillator system. The machine-learned model is a high-dimensional dynamical system that displays not only the trained dynamical behaviors of the true oscillator (i.e., attractor reconstruction), but also the untrained networked dynamical behaviors of the true oscillator.

Future work includes extending the network stability estimation technique for synchronization on networks with delays \cite{dhamala2004enhancement,flunkert2010synchronizing}, nonlinear coupling (known or unknown) \cite{stankovski2017coupling}, time-varying connections \cite{sorrentino2008adaptive,ghosh2022synchronized} and cluster synchronization \cite{pecora2014cluster,hart2019topological}, and for stability metrics other than the largest Lyapunov exponent \cite{pecora1998master,illing2002synchronization}. The MSF formalism only tells about local stability; It would also be interesting to test whether the trained reservoir could be used to predict the basin stability \cite{menck2013basin} as well. For cluster synchronization and basin stability, the reservoir would need to be able to model dynamics away from the attractor on which it was trained. Recent developments \cite{rohm2021model,kong2023reservoir} may enable this.

\textbf{Acknowledgments}:  JDH acknowledges support from the Office of the Secretary of Defense through the Applied Research for Advancement of S\&T Priorities (ARAP) program under the Neuropipe
project.

%\end{thebibliography}
\bibliographystyle{unsrt}
\bibliography{sample}

\end{document}

\renewcommand{\theequation}{S.\arabic{equation}}
\renewcommand{\thefigure}{S.\arabic{figure}}
\title{Supplemental Material for \\ ``Estimating the master stability function from the time series of one oscillator via reservoir computing''}
\author{Joseph D. Hart \textsuperscript{*}}
%\affiliation{
 \affil{
 US Naval Research Laboratory, Code 5675, Washington, DC, USA 20375 \\ *joseph.hart@nrl.navy.mil
}
%\email{joseph.hart@nrl.navy.mil}

\maketitle

\section{Justification for the form of the coupling}
In this section, we present a justification for why one might expect that the form of the coupling between reservoir computers (RCs) described by Eq. 6 is a good model for the true coupling described by Eq. 2. 

We begin by recalling that what we need are \textit{A)} the dynamics of the flow on the synchronization manifold (which are given by the trained reservoir Eq. 5 and \textit{B)} the variational equation given by

\begin{equation}
\label{eq:flow_variational}
    \delta\dot{\mathbf{x}}_i(t) = d\mathbf{F}(\mathbf{x}_s(t)) \delta\mathbf{x}_i(t)- \epsilon L_{ij} d\mathbf{H}(\mathbf{x}_s(t))\delta\mathbf{x}_j(t).
\end{equation}

A RC trained on a single node is described by Eqs. 4 and 5. Let us define a function $\mathbf{G}$ such that $\hat{\mathbf{x}}[t]\equiv \mathbf{G}(\hat{\mathbf{x}}[t-1])$. From Eq. 4, we have $\mathbf{G}(\hat{\mathbf{x}}[t])=W^{out}\mathbf{r}[t]$. The function $\mathbf{G}$ describes what the RC has learned: a mapping from $\hat{\mathbf{x}}[t-1]$ to $\hat{\mathbf{x}}[t]$. If we consider small $\tau$, we can approximate the uncoupled flow from Eq. 1 as $\mathbf{x}[t]\approx \mathbf{x}[t-1]+\tau\mathbf{F}(\mathbf{x}[t-1])$, which apparently implies that 
\begin{equation}
    \mathbf{G}(\hat{\mathbf{x}}[t])\approx \mathbf{x}[t]+\tau\mathbf{F}(\mathbf{x}[t]) %+ \mathcal{O}(\tau^2).
\end{equation}
Therefore, if we examine the function $\mathbf{G}$ applied to $\hat{\mathbf{x}}_i[t]-\tau\epsilon L_{ij}\mathbf{H}(\hat{\mathbf{x}}_j[t]))$, we find  
\begin{equation}
\label{eq:G2}
    \mathbf{G}(\hat{\mathbf{x}}_i[t]-\tau\epsilon L_{ij}\mathbf{H}(\hat{\mathbf{x}}_j[t]))\approx \mathbf{x}_i[t]+\tau\{\mathbf{F}(\mathbf{x}_i[t]-\tau\epsilon L_{ij}\mathbf{H}(\mathbf{x}_j[t])) - \epsilon L_{ij}\mathbf{H}(\mathbf{x}_j[t])\}.
\end{equation}
We note that Eq. \ref{eq:G2} is not the same as the right-hand-side of the Euler discretization of Eq. 2. If Eq. \ref{eq:G2} is Taylor expanded about small $\tau$ and terms of $\mathcal{O}(\tau^2)$ are discarded, then one does obtain  the Euler discretization of Eq. 2; however, in order to discard the term of $\mathcal{O}(\tau^2)$, it must be the case that $|\tau\mathbf{F}^\prime(\mathbf{x}_i)\epsilon L_{ij}\mathbf{H}(\mathbf{x}_j)|\ll|\mathbf{F}(\mathbf{x}_i)|$. This is generally not a valid assumption for the values of $\tau$ used here because $\sum_jL_{ij}\mathbf{H}(\mathbf{x}_j)$ can be large. However, when the network is synchronized $\sum_jL_{ij}\mathbf{H}(\mathbf{x}_s)=0$ so $|\tau\mathbf{F}^\prime(\mathbf{x}_s)\epsilon L_{ij}\mathbf{H}(\mathbf{x}_s)|=0$. Motivated by this and by the fact that for computing the MSF we need only the variational equation, we Taylor expand Eq. \ref{eq:G2} about the synchronization manifold, resulting in

\begin{equation}
\begin{split}
    d\mathbf{G}(\hat{\mathbf{x}}_s[t])&(\delta\mathbf{x}_i(t) -\tau\epsilon L_{ij}d\mathbf{H}(\hat{\mathbf{x}}_s[t]) \delta\mathbf{x}_j[t]) \\ &\approx \delta\mathbf{x}_i[t]+\tau\{d\mathbf{F}(\mathbf{x}_s[t])(\delta\mathbf{x}_i[t]- \tau\epsilon L_{ij}d\mathbf{H}(\mathbf{x}_s[t])\delta\mathbf{x}_j[t] ) - \epsilon L_{ij}d\mathbf{H}(\mathbf{x}_s[t])\delta\mathbf{x}_j[t]\},
    \end{split}
\end{equation}
which reduces to the Euler approximation of Eq. \ref{eq:flow_variational} if it is assumed that $|\tau\epsilon L_{ij}d\mathbf{H}(\mathbf{x}_s[t])\delta\mathbf{x}_j[t]|\ll|\delta\mathbf{x}_i[t]|$ (that is, if one can ignore the term $\tau\epsilon L_{ij}d\mathbf{H}(\mathbf{x}_s[t])\delta\mathbf{x}_j[t]$). This means that, near the synchronization manifold, one might expect the stability of synchronization of well-trained RCs coupled as in Eq. 6 to be similar to that of the true coupled oscillator system in Eq. 2 as long as this approximation is satisfied. We now investigate this approximation.

If we transform to the basis in which the Laplacian matrix is diagonal, we obtain

\begin{equation}
\label{eq:}
\begin{split}
    d\mathbf{G}(\hat{\mathbf{x}}_s[t])&(\delta\mathbf{q}_i(t) -\tau\epsilon \lambda_i d\mathbf{H}(\mathbf{\hat{x}}_s[t]) \delta\mathbf{q}_i(t)) \\& \approx \delta\mathbf{q}_i[t]+\tau\{d\mathbf{F}(\mathbf{x}_s[t])(\delta\mathbf{q}_i[t]- \tau\epsilon \lambda_i d\mathbf{H}(\mathbf{x}_s[t])\delta\mathbf{q}_i[t] ) - \epsilon \lambda_i d\mathbf{H}(\mathbf{x}_s[t])\delta\mathbf{q}_i[t]\}.
    \end{split}
\end{equation}
It is now clear that  $|\tau\epsilon L_{ij}d\mathbf{H}(\mathbf{x}_s[t])\delta\mathbf{x}_j[t]|\ll|\delta\mathbf{x}_i[t]|$ is equivalent to $|\tau\epsilon \lambda_i d\mathbf{H}(\mathbf{x}_s(t))\delta\mathbf{q}_i[t]|\ll|\delta\mathbf{q}_i[t]|,$ which is easier to interpret. Indeed, in the case of linear coupling considered in this manuscript, this can be further simplified to the assumption that 
\begin{equation}
\label{eq:validity}
    \tau\ll\frac{1}{\epsilon\lambda}\equiv\frac{1}{K},
\end{equation} where $K$ is the normalized coupling strength. Equation \ref{eq:validity} gives an approximate domain of validity of our MSF computation, and suggests that the RC approximation of the MSF should be more accurate for smaller $K$. Indeed, this is what we observe in Fig. 1 of the main text.

An important takeaway from this section is that the factor of $\tau$ in the RC coupling term is essential to obtain an accurate estimate of the MSF. While coupling RCs has been studied extensively \cite{weng2019synchronization,fan2021anticipating,hu2022synchronization,weng2023synchronization}, as far as we are aware, this is the first time that this factor of $\tau$ has been used to link the behavior of coupled RCs to the behavior of coupled true oscillators.

\section*{The H{\'e}non Map}
In order to demonstrate the generality of our approach and to facilitate some analysis,%and to try to understand whether the true MSF can be well-approximated by a RC trained on only one of the dynamical variables (rather than on all of the dynamical variables in the case of the Lorenz system described above),
the H{\'e}non map is considered in this section. The H{\'e}non map is described by

\begin{equation}
    \begin{bmatrix}
    x[n+1] \\ y[n+1]
    \end{bmatrix} = \mathbf{F}\Bigg(\begin{bmatrix}
    x[n] \\ y[n]
    \end{bmatrix}\Bigg) =
    \begin{bmatrix}
     1 + y[n] - ax^2[n] \\
     bx[n]
     \end{bmatrix}
\end{equation}
Typically, the values for $a$ and $b$ are taken to be 1.4 and 0.3, respectively, and we use these values here. With these parameters, the H{\'e}non map is chaotic with Lyapunov exponents 0.42 and -1.6. 

We consider a network of linearly-coupled H{\'e}non maps, described by

\begin{equation}
\label{eq:coupledHenon}
    \mathbf{x}_i[n+1] =
   \mathbf{F}\big(\mathbf{x}_i-\epsilon L_{ij}E\mathbf{x}_j[n]\big),
\end{equation}
where $\mathbf{x}[n]=(x[n],y[n])^T$ and
\begin{equation*}
E=\begin{bmatrix}
    1 & 0 \\ 0 & 0
\end{bmatrix}.
\end{equation*}
The true master stability equation is given by 
\begin{equation}
    \delta \mathbf{y}[n+1] = d\mathbf{F}(\mathbf{x}_s[n])(\mathbbold{1}_2-K E)\delta \mathbf{y}[n],
\end{equation}
and the true MSF is given by the largest Lyapunov exponent of this equation as a function of the normalized coupling strength $K$. 

\begin{figure}[h!]
    \centering
  \begin{subfigure}{0.48\textwidth}
\includegraphics[width=\textwidth]{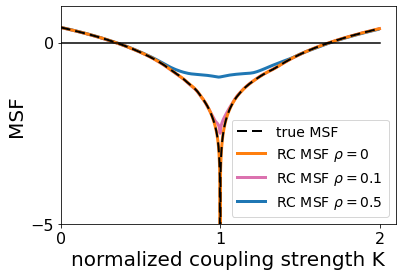}
  \end{subfigure}
  \begin{subfigure}{0.48\textwidth}
\includegraphics[width=\textwidth]{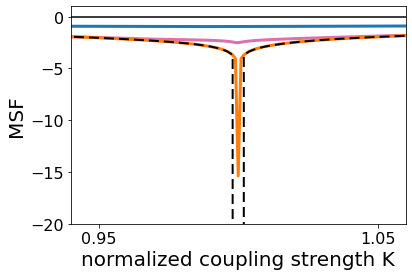}
  \end{subfigure}
  
    \caption{\textit{a}) Comparison of the MSF of the H{\'e}non map with the MSF of a RC trained on both variables of the H{\'e}non map with different values of the spectral radius of the reservoir adjacency matrix. \textit{b}) Zoom of \textit{a}. The true MSF is $-\infty$ at K=1.}
    \label{fig:HenonMSF}
\end{figure}

Initially, a RC with 40 nodes, spectral radius 0.1, $\sigma=0.05$, and average in-degree of 6 was used.  Here, we use $\mathbf{P}(\mathbf{x})=\mathbf{x}$, $\rho=0.1$ and $\sigma=0.1$. $W^{in}$ was chosen in the same way as was used in the main text for the Lorenz system. Each bias $b_i$ was chosen randomly from an i.i.d. uniform distribution with range [-1,1]. For training, the reservoir was driven by the $x$ and $y$ variables of the H{\'e}non map, and 8000 steps were used for training. When operating in feedback mode, the two largest Lyapunov exponents of the RC were 0.42 and -1.6, which are in agreement with the known true values. The MSF of this trained RC was computed (i.e., the largest Lyapunov exponent of Eq. 8 with $\tau=1$). The results are shown in Fig. \ref{fig:HenonMSF}. The H{\'e}non MSF and the RC MSF generally agree quite well, except for when $K\to 1$. In this case, the true MSF $\to -\infty$, while the RC estimate of the MSF is negative but finite. 

The reason for this disagreement is the following. The true MSF is the largest Lyapunov exponent of the variational equation:

\begin{equation}
\label{eq:variationalHenon2}
\begin{split}
    \delta \mathbf{y}[n+1] &= \begin{bmatrix}
        -2ax & 1 \\ b & 0
    \end{bmatrix}\cdot \Bigg(\begin{bmatrix}
        1 & 0 \\ 0 & 1
    \end{bmatrix}-K\begin{bmatrix}
        1 & 0 \\ 0 & 0
    \end{bmatrix}\Bigg)\delta \mathbf{y}[n] \\ &= \begin{bmatrix}
        -2ax(1-K) & 0 \\ 0 & 0
    \end{bmatrix}\delta \mathbf{y}[n]= J\delta \mathbf{y}[n].
    \end{split}
\end{equation}    
When $K\to 1$, The Jacobian matrix $J\to 0$, and the Lyapunov exponents of Eq. \ref{eq:variationalHenon2} approach $-\infty$. However, the reservoir Jacobian matrix has a different limit:

\begin{equation}
\begin{split}
J_{ij}&=\mathrm{sech}^2\big(B_{ip}r^{(s)}_p[t]+b_i\big)\{A_{ij} + W_{ik}^{in}W_{kj}^{out} - \tau K W_{ik}^{in}E_{k\ell}W_{\ell j}^{out}\}\\
&\to\mathrm{sech}^2\big(B_{ip}r^{(s)}_p[t]+b_i\big)\{A_{ij} + W_{ik}^{in}W_{kj}^{out} - \tau W_{i1}^{in}W_{1 j}^{out}\},
\end{split}
\end{equation}
which is non-zero in general, so the Lyapunov exponents are finite. In the special case where $E=\mathbbold{1}$ (and $K=1$), the reservoir Jacobian limits to
\begin{equation}
     J_{ij}\to\mathrm{sech}^2\big(B_{ip}r^{(s)}_p[t]+b_i\big)A_{ij}.
\end{equation}
Even in this special case, the reservoir Lyapunov exponents are finite because of the presence of the reservoir adjacency matrix.

In order to test this, the H{\'e}non map (with $E$ given by Eq. \ref{eq:coupledHenon}) was modeled with a ``RC'' with no adjacency matrix (spectral radius equal to 0).  This is often called an ``extreme learning machine'' \cite{huang2006extreme}. The MSF from this RC is shown in orange in Fig. \ref{fig:HenonMSF}. The agreement is somewhat better, but the MSF of the RC is about -15 at $K=1$.

\begin{figure}[h!]
    \centering
    \includegraphics[width=0.5\textwidth]{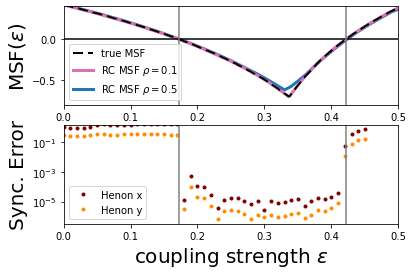}
    \caption{\textit{Top} Comparison of the MSF of the H{\'e}non map (black dotted line) with the MSF of RCs with different spectral radii trained on both variables of the H{\'e}non map. Here the MSFs are given as a function of coupling strength $\epsilon$ rather than of normalized coupling strength $K$. \textit{Bottom} Synchronization error for a network of coupled H{\'e}non maps with the adjacency matrix given by Eq. \ref{eq:HenonLmat}. The solid gray line gives the prediction of the stability boundary by the MSF of the RC trained on the H{\'e}non map.}
    \label{fig:HenonSE}
\end{figure}

For comparison, we also computed the MSF for a RC with spectral radius = 0.5; the MSF from this RC is shown in blue in Fig. \ref{fig:HenonMSF}, and displays the worst accuracy near K=1. This confirms that better RC estimates of the MSF can be obtained for RCs with smaller $\rho$ (assuming, of course, that good attractor reconstruction or at least short-term prediction is still attainable). We found that to be the case for the Lorenz and R{\"o}ssler systems as well.

It is now clear that the MSF of the trained RC cannot perfectly equal the MSF of the H{\'e}non map. Nevertheless, we find that the MSF of the trained RC provides a sufficiently accurate estimation of the MSF to provide useful information about the stability of synchronization of a given network; in other words, the sign of the reservoir MSF seems to be an accurate approximation of the sign of the true H{\'e}non MSF, even when the values of the MSF differ. Since it is the sign of the MSF that ultimately determines the stability, the RC estimate of the MSF is useful.

Further, when the magnitude of the convergence to synchronization is important, the value that determines the stability of a network of oscillators is  $\max_i(MSF(K_i=\epsilon\lambda_i))$, where $\lambda_i$ are the eigenvalues of the Laplacian coupling matrix. The $-\infty$ value will not be the maximum, so we do not anticipate this being a problem in practice.

For confirmation of these results, a network of four coupled H{\'e}non maps were directly simulated using Eq. \ref{eq:coupledHenon} for a variety of $\epsilon$ with 

\begin{equation}
\label{eq:HenonLmat}
  L=\begin{bmatrix}
      2&-1&0&-1\\
      -1&2&-1&0\\
      0&-1&2&-1\\
      -1&0&-1&2
  \end{bmatrix},  
\end{equation}
which has eigenvalues 0, 2, and 4. The network synchronization error and the MSFs (true and RC) as a function of coupling strength $\epsilon$ are shown in Fig. \ref{fig:HenonSE}. The agreement of the synchronization error with the stability boundary predicted by both MSF is excellent. We also see that the agreement between the reservoir MSF and the true MSF are now excellent, even for the case when $\rho=0.5$, because the MSF as a function of $\epsilon$, which is defined as $MSF(\epsilon)=\mathrm{max}_i(MSF(\epsilon\lambda_i))$, is now plotted. The $-\infty$ Lyapunov exponent is never the maximum.

\section*{Supplemental Material: The R{\"o}ssler System}

In this section, we demonstrate that our MSF estimation method can also provide a good estimate of the true MSF even when a reliable reproduction of the target system dynamics can not be obtained. To demonstrate this, we use the R{\"o}ssler system, for which it has been found that obtaining a reliable attractor reconstruction using reservoir computing is more difficult than for the Lorenz system \cite{scully2021measuring}. We considered a R{\"o}ssler system with the same equations and parameters as in Ref. \cite{huang2009generic}.

\begin{figure}[h!]
    \begin{subfigure}{0.45\textwidth}    \includegraphics[width=\textwidth]{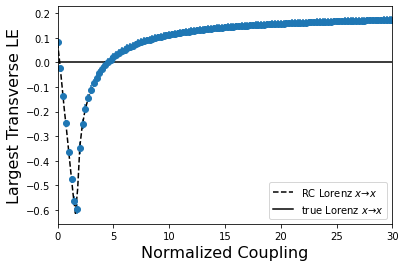}
    \caption{}
    \end{subfigure}
    \begin{subfigure}{0.45\textwidth}    \includegraphics[width=\textwidth]{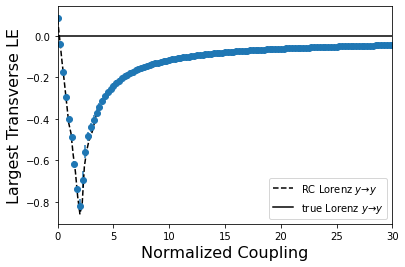}
    \caption{}
    \end{subfigure}
    \caption{Comparison of the MSF of a trained RC driven by measured data with the MSF of the true R{\"o}ssler system on which the RC was trained for linear coupling of (\textit{a}) $x\to x$ and (\textit{b}) $y\to y$.}
    \label{fig:rossler}
\end{figure}

As for the Lorenz system, we use a 300 node RC consisting of an Erd{\"o}s-Renyi network of $tanh$ nonlinearities. Here, we use $\mathbf{P}(\mathbf{x})=\mathbf{x}$, $\rho=0.1$ and $\sigma=0.1$. $W^{in}$ was chosen in the same way as was used in the main text for the Lorenz system. Each bias $b_i$ was chosen randomly from an i.i.d. uniform distribution with range [-1,1].

We did not find that the RC gave a good reconstruction of the R{\"o}ssler attractor. Therefore, in place of $\mathbf{\hat{x}}_s$ generated by the RC, we used the measured values of the true system (the same values as were used to train the RC). The results for a single RC are shown in Fig. \ref{fig:rossler}. We estimated only the MSFs for $x\to x$ and $y\to y$ coupling because these are the only single-variable, linear coupling schemes that lead to global synchronization \cite{huang2009generic}. The agreement of the MSF of the RC with the MSF of the true R{\"o}ssler system is excellent.

\bibliographystyle{unsrt}
%\bibliography{sample}